\documentclass[11pt,a4paper,english,superscriptaddress,aps]{revtex4}

\usepackage{amsmath}
\usepackage{amssymb}
\makeatletter
\usepackage{babel}
\usepackage{hyperref}

\begin{document}

\title{Scattering of spin $1/2$ particles by the $2+1$ dimensional noncommutative Aharonov-Bohm potential}

\author{A. F. Ferrari} 
\author{M. Gomes} 
\author{C. A. Stechhahn}

\affiliation{Instituto de F\'\i sica, Universidade de S\~ao Paulo\\
Caixa Postal 66318, 05315-970, S\~ao Paulo, SP, Brazil}

\email{alysson, mgomes, carlos@fma.if.usp.br}

\begin{abstract}
In this work we study modifications in the Aharonov-Bohm effect for relativistic spin $1/2$ particles due to the noncommutativity of spacetime in $2 + 1$ dimensions. The noncommutativity gives rise to a correction to the Aharonov-Bohm potential which is highly singular at the origin, producing divergences in a perturbative expansion around the usual solution of the free Dirac equation. This problem is surmounted by using a perturbative expansion around the exact solution of the \textit{commutative} Aharonov-Bohm problem. We calculate, in this setting, the scattering amplitude and the corrections to the differential and total cross sections for a spin $1/2$ particle, in the small-flux limit.
\end{abstract}

\maketitle

The idea of the noncommutativity of spacetime became quite popular in the last years as a proposal for a more fundamental description of the spacetime at the Planck scale~\cite{Doplicher:1994tu} or, alternatively, as a particular limit of a quantum gravity theory~\cite{Seiberg4}. A great deal of literature have been devoted to the exploration of the consequences of this noncommutativity in several field theoretical and quantum mechanical models (we refer to the reviews in~\cite{reviews,szabo} for an extensive list of references). On a more general perspective, one faces the noncommutativity as a particular sort of deformation, involving nonlocal interactions of a very specific nature, that can lead to peculiar effects. From this viewpoint, it is interesting to examine the behavior of different models when they are supposed to live in a noncommutative spacetime.

In the so-called canonical noncommutativity, spacetime coordinates $q^\mu$ are supposed to satisfy
\begin{equation} \label{eq1.1}
[{q}_\mu,{q}_\nu]=i\theta_{\mu\nu}\,,
\end{equation}

\noindent
where $\theta_{\mu\nu}$ is a constant anti-symmetric matrix of dimension length squared. 
Instead of dealing with quantum operators $\Phi (q^\mu)$ one can work with functions of the commutative variables $x^\mu$ endowed with the noncommutative Moyal-Groenewold product (see~\cite{szabo}, for example), 
\begin{equation} \label{produtoMoyal}
\phi_{1}(x)*\phi_{2}(x)=\lim_{y\rightarrow x}e^{\frac{i}{2}\theta^{\mu
\nu}\frac{\partial}{\partial y^\mu}\frac{\partial}{\partial x^\mu}}\phi_{1}(y)\phi_{2}(x)\,.
\end{equation}

\noindent
Such a non-commutative algebra defines a quantum field theory by means of functional integration, which leads to a perturbative scheme very similar to the usual (commutative) one. Several interesting effects of the underlying noncommutativity of spacetime have been explored this way, such as the ultraviolet/infrared (UV/IR) mixing~\cite{Minwalla5}. Except for some supersymmetric theories~\cite{wesszumino,Mat,Zanon:2000nq,Bichl:2002wb}, or in some alternative formulations~\cite{Balachandran:2006pi,Smailagic:2004yy,Anacleto:2005mq,Charneski,Anacleto:2006eu}, the presence of UV/IR mixing ruins most of the usual perturbative schemes. Others prominent features of this approach to the noncommutative spacetime is the violation of unitarity and causality when $\theta_{0i}\neq 0$ (see~\cite{foot1}), as well as strong restrictions on gauge groups and couplings in gauge theories~\cite{Matsubara:2000gr,Chaichian,Ferrari2}. The effects of the spacetime noncommutativity has also been studied in the context of non-relativistic quantum mechanics, see f.e.~\cite{Gamboa:2000yq,Nair:2000ii,Lubo:2003bm,Girotti1,Bemfica:2005pz}.

In this work, we examine the effects of the noncommutativity of spacetime in the scattering of spin $1/2$ particles from an Aharonov-Bohm (AB) potential. The AB effect~\cite{Aharonov10}, one of the most interesting problems of planar quantum dynamics, is related to the scattering of charged particles by an impenetrable solenoid of small radius. In the noncommutative situation the AB effect has been studied in the context of quantum mechanics~\cite{Sheikh-Jabbari16,Mendez18,Li:2005mi} as well as in the quantum field theory approach~\cite{Anacleto:2004qq,Anacleto:2004xs}. 

We consider the noncommutative plane with a thin solenoid perpendicular to it. The coordinates $x^i$ satisfy the commutation relation in Eq.~(\ref{eq1.1}) with $\theta^{ij}=\theta\epsilon^{ij}$, $\epsilon^{ij}$ being the Levi-Civit\`a symbol, normalized according to $\epsilon^{12}=1$. 
A magnetic field is supposed to exist only inside the solenoid, which can be derived from a potential $A_k$ that has been already calculated in the noncommutative case in~\cite{Mendez18}, 
\begin{eqnarray} \label{Ak}
A_{k}=-\epsilon_{kl}x_l\left(\frac{\Phi}{2\pi}\frac{1}{r^2}-\theta\frac{\Phi^2}{8\pi^2r^4}\right)\, + \cdots\,,
\end{eqnarray}

\noindent 
where the dots stand for terms of higher order in the noncommutative parameter $\theta$, or $\Phi$, the magnetic flux through the plane. In this work we use natural units, so that $c=\hbar=1$, and our spacetime metric is $\eta^{\mu \nu} = {\rm diag}(1,-1,-1)$. The Dirac equation that describes relativistic spin $1/2$ particles in the AB potential in the NC case is,
\begin{equation}  \label{diraceq}
[i\gamma^\mu(\partial_\mu-ieA_\mu)-m]*\psi=0
\end{equation}

\noindent 
involving the Moyal product defined by Eq.~(\ref{produtoMoyal}). Since a theory with two-component spinors in three dimensions violates parity, we have to choose to work with either spin $s=+1/2$ or $s=-1/2$ particles; in this work, we focus on the $s=+1/2$ case, but our results hold for both choices of the spin. The Dirac matrices, in this case, are given in terms of the usual Pauli matrices as $(\gamma^0, \gamma^1, \gamma^2) = (\sigma_3, i \sigma_1, i\sigma_2)$, and one can readily rewrite Eq.~(\ref{diraceq}) as, 
\begin{eqnarray} 
\label{EqDiracPotlABNCcomAk}
i\frac{\partial\psi}{\partial t}=\left(-i\gamma^0\gamma^l\partial_l+m\gamma^0\right)\psi-e\left(\gamma^0\gamma^{k}A_{k}+\frac{i\theta^{ij}\gamma^{0}\partial_{i}(\vec{\gamma}\cdot\vec{A})\partial_j}{2}\right)\psi\,,
\end{eqnarray}

\noindent 
from which follows the Hamiltonian $H=H_0 + H_{\rm int}$, where
\begin{equation}
H_{0}=-i\gamma^0\gamma^{\ell}\partial_{\ell}+m\gamma^0
\end{equation}

\noindent 
and
\begin{eqnarray} 
H_{\rm int}&=&e\frac{\Phi}{2\pi}\frac{\gamma^0\gamma^{k}\epsilon_{k\ell}x_{\ell}}{r^2}
\nonumber
\\
&-&\theta\left[e\gamma^{0}\gamma^{k}\epsilon_{k\ell}x_\ell
\frac{\Phi^2}{8\pi^2r^4}-\frac{ie}{2}\frac{\Phi}{2\pi}\frac{\gamma^{0}\gamma^{j}}{r^2}\partial_j+ie\frac{\Phi}{2\pi}\frac{\gamma^{0}\gamma^{i}x_ix_j}{r^4}\partial_j\right] \,,
\end{eqnarray}

\noindent
up to order $\theta$. From now on, we also disregard terms of order $\Phi^2$, so that in our calculations $H_{\rm int}$ reduces to
\begin{equation} \label{HintNC1}
H_{\rm int}\,=\,e\frac{\Phi}{2\pi}\frac{\gamma^0\gamma^{k}\epsilon_{k\ell}x_{\ell}}{r^2}
-\theta\left[
ie\frac{\Phi}{2\pi}\frac{\gamma^{0}\gamma^{i}x_ix_j}{r^4}\partial_j -\frac{ie}{2}\frac{\Phi}{2\pi}\frac{\gamma^{0}\gamma^{j}}{r^2}\partial_j
\right]\,.
\end{equation}

The Hamiltonian in Eq.~(\ref{HintNC1}) is very singular at the origin. In order to avoid divergences in the intermediate calculations we regularize the interaction Hamiltonian introducing a parameter $\lambda$ such that,
\begin{equation}
H_{\rm int}=\lim_{\lambda\rightarrow 1}H_{{\rm int},\lambda}\,,
\end{equation}

\noindent 
where $H_{{\rm int},\lambda}$ is given by, 
\begin{equation} \label{Hdelambda}
H_{{\rm int},\lambda} = e\frac{\Phi}{2\pi} \frac{\gamma^0\gamma^{k} \epsilon_{k\ell}x_{\ell}}{r^{2\lambda}} - \theta\left[ie\frac{\Phi}{2\pi}\frac{\gamma^{0}\gamma^{i}x_ix_j}{r^{2\lambda+2}}\partial_j
-\frac{ie}{2}\frac{\Phi}{2\pi}\frac{\gamma^{0}\gamma^{j}}{r^{2\lambda}}\partial_j
\right]\,.
\end{equation}

\noindent 
For $\lambda < 1$, the regularized Hamiltonian $H_{{\rm int},\lambda}$ does not give rise to any divergence. After all integrations are performed, we will investigate whether the ``physical'' limit $\lambda \rightarrow 1$ can be taken. In the Born approximation, one obtains for the scattering amplitude of this problem the expression 
\begin{equation} \label{SfiHintNC}
T_{fi}=-i\frac{\delta(E_f-E_i)}{2\pi}\lim_{\lambda\rightarrow1}\int d^{2}x\, e^{i\vec{k'}\cdot\vec{x}}\,u^{\dagger}(\vec{k'}) \, H_{{\rm int},\lambda} \, u(\vec{k})e^{-i\vec{k}\cdot\vec{x}}\,,
\end{equation}

\noindent 
if the limit exists. Here, $u(k)$ is the solution of the free Dirac equation, 
\begin{equation}
u(k)\,=\,\frac{1}{\sqrt{2E}} \left[ \begin{matrix} \sqrt{E+m} \\ \sqrt{E-m} \, (-i)e^{i\xi} \end{matrix} \right] \frac{e^{i\vec{k}\cdot \vec{x} } }{2\pi}\,.
\end{equation}

\noindent 
The problem with such a procedure is that we find a divergence when removing the regularization, after performing the space integral in Eq.~(\ref{SfiHintNC}). One possible solution to this problem, which we will follow in this work, is to adopt a different perturbation scheme. Suppose that we can split $H_{{\rm int},\lambda} = H_{{\rm int},\lambda}^{(1)}+H_{{\rm int},\lambda}^{(2)}$ such that we can find the explicit eigenfunctions $\psi^{(1)}_\lambda$ of $H_0 + H_{{\rm int},\lambda}^{(1)}$. In this case, a Born-like approximation leads to~\cite{Messiah}
\begin{equation}  \label{Born1}
T_{fi}\,=\,\lim_{\lambda\rightarrow1}\left(T_{fi}^{(1)} + \,T_{fi}^{(2)}\right)_{\lambda} = \,\lim_{\lambda\rightarrow1}T_{fi,\lambda}^{(1)} - i \lim_{\lambda\rightarrow1} \int dt \, \left\langle \psi^{(1)}_{f,\lambda} | H_{{\rm int},\lambda}^{(2)} | \psi^{(1)}_{i,\lambda} \right\rangle \,,
\end{equation}

\noindent
where $T_{fi,\lambda}^{(1)}$ is the transition amplitude from $\psi^{(1)}_{i,\lambda}$ to $\psi^{(1)}_{f,\lambda}$. By a clever choice of $H_0 + H_{{\rm int},\lambda}^{(1)}$, we will show that the aforementioned difficulty does not appear, and the $\lambda \rightarrow 1$ limit in Eq.~(\ref{Born1}) can be safely performed.  

A natural choice for $H_0 + H_{{\rm int}}^{(1)}$ is the total Hamiltonian of the commutative Aharonov-Bohm case,
\begin{equation}
\label{hin1}
H_0 + H_{{\rm int}}^{(1)} = -i\gamma^0\gamma^{\ell} \partial_{\ell}+m\gamma^0 + e\frac{\Phi}{2\pi}\frac{\gamma^0\gamma^{k}\epsilon_{k\ell}x_{\ell}}{r^2}\,.
\end{equation}

\noindent
(notice that we omitted the $\lambda$ dependence in $H_{\rm int}^{(1)}$ since this Hamiltonian does not induce singularities, hence we just take $\lambda \rightarrow 1$ for this term once and for all). As it has been discussed elsewhere~\cite{Alford:1988sj,deSousaGerbert:1988yt,Hagen11,Girotti:1996jw,Coutinho:1993cp,Coutinho:1994xk}, the Hamiltonian~(\ref{hin1}) has an infinite number of self-adjoint extensions, depending on the boundary conditions imposed at the origin. Studying the Aharonov-Bohm problem as a limit of certain radially symmetric magnetic configurations, two solutions are singled out. They are characterized by the fact that only the upper (lower) component of the spinor has a singularity at the origin, and they correspond to systems that are identical except for the sign of the magnetic flux~\cite{deSousaGerbert:1988yt}. Here, we choose to work with eigenfunctions whose upper components are regular at the origin. Other self-adjoint extensions can in principle be similarly treated.

The solution of the commutative Aharonov-Bohm problem that we will use can be cast as~\cite{Girotti:1996jw},
\begin{equation} \label{eqprincipalFinal}
\psi_{k,\xi}^{(1)}(\vec{r})=\frac{1}{(2\pi)\sqrt{2E}}\sum_{\ell=-\infty}^{\infty}
C_\ell 
{\left[
\begin{matrix}\sqrt{E+m}J_{\mid\ell-\alpha\mid}(kr)\\
\sqrt{E-m}e^{i\varphi}\epsilon(\ell-\alpha)J_{\mid\ell-\alpha\mid+\epsilon(\ell-\alpha)}(kr)\end{matrix}
\right]e^{i\ell(\varphi-\xi)}}\,,
\end{equation}

\noindent 
where
\begin{equation} \label{CoeficienteCell}
C_\ell=(-i)^{|\ell-\alpha|}(-1)^{\ell}\,,
\end{equation}
\begin{equation}
\alpha = {e \Phi}/{2\pi}\,,
\end{equation}

\noindent 
$\xi$ is the angle between the incident flux of particles with momentum $k$ and the $x_1$ axis, $\epsilon(x) = 1$ for $x \geq 0$ and $\epsilon(x) = -1$ for $x<0$, and $J_n$ are Bessel's functions of the first kind~\cite{watson}. Unless for the $s$-wave, which for small $r$ behaves as $r^{-\alpha}$, all other terms in the sum~(\ref{eqprincipalFinal}) vanish at the origin, so one might guess that $\psi_{k,\xi}^{(1)}(\vec{r})$ is a good candidate to be used in Eq.~(\ref{Born1}).  As we shall prove, the $s$-wave singularity is not harmful insofar the regularization is removed after integration. This choice of $\psi^{(1)}$ will be effective in controlling the small-$r$ singularities of the Hamiltonian in Eq.~(\ref{Hdelambda}). 
Clearly, in our case, the part of the interaction Hamiltonian to be treated perturbatively is the one containing the noncommutativity parameter $\theta$, 
\begin{eqnarray}
H_{{\rm int},\lambda}^{(2)} = -\theta\left[-\frac{ie}{2} \frac{\Phi}{2\pi}\frac{\gamma^{0}\gamma^{j}}{r^{2\lambda}} \partial_j+ie\frac{\Phi}{2\pi}\frac{\gamma^{0}\gamma^{i}x_ix_j}{r^{2\lambda+2}}\partial_j\right]\,.
\end{eqnarray}

To simplify, we choose $\xi = 0$, so that the incident flux is parallel to the negative $x_1$ axis. In this situation, Eq.~(\ref{eqprincipalFinal}) can be cast as
\begin{equation}
\psi_{k}^{(1)}(\vec{r})=\sum_\ell C_\ell\left[\begin{matrix}F_1(r,\varphi)\cr
F_2(r,\varphi)\end{matrix}\right]\,,
\end{equation}

\noindent 
where
\begin{equation} \label{F1}
F_1(r,\varphi)=e^{i\ell\varphi}J_{|\ell-\alpha|}(kr)\frac{\sqrt{E+m}}{(2\pi)\sqrt{2E}}\,,
\end{equation}

\noindent 
and
\begin{equation} \label{F2}
F_2(r,\varphi)=e^{i(\ell+1)\varphi}\epsilon(\ell-\alpha)J_{|\ell-\alpha|+\epsilon(\ell-\alpha)}(kr)\frac{\sqrt{E-m}}{(2\pi)\sqrt{2E}}\,.
\end{equation}

\noindent
The product $ (\psi_{\hat k}^{(1)})^\dagger  H_{{\rm int},\lambda}^{(2)}  \psi_{k}^{(1)} $ is written as
\begin{eqnarray} 
\label{EqLimLamb1DerivEmF1eF2}
(\psi_{\hat k}^{(1)})^\dagger  H_{{\rm int},\lambda}^{(2)}  \psi_{k}^{(1)} & = & -b\theta \, \sum_{\hat \ell}\sum_{\ell}C_{\hat \ell}^{\dagger} C_{\ell}\left[\begin{array}{cc}{\hat F}_1^\dagger (r,\varphi)
  &
  {\hat F}_2^\dagger (r,\varphi) \end{array} \right] \left(-\frac{\gamma^0\gamma^j}{2r^{2\lambda}}\partial_j + \frac{\gamma^0\gamma^{i}x_ix_j}{r^{2\lambda+2}} \partial_j\right)\times
\nonumber\\
&\times&
\left[\begin{matrix}F_1(r,\varphi)\\F_2(r,\varphi)\end{matrix}
\right]\,,
\end{eqnarray}

\noindent 
where $b={\Phi}/{2\pi}$. Hereafter, all hatted quantities refer to the final state of the scattering. Using the explicit expressions in Eqs.~(\ref{F1}) and~(\ref{F2}), one can write,
\begin{equation}
(\psi_{\hat k}^{(1)})^\dagger  H_{{\rm int},\lambda}^{(2)}  \psi_{k}^{(1)} = -b \theta \, \sum_{\hat \ell} \sum_{\ell}C_{\hat \ell}^{\dagger}C_{\ell}\left[\begin{array}{cc}{\hat F}_1^{\dagger}(r,\varphi)
      & {\hat F}_2^{\dagger}(r,\varphi) \end{array} \right] \left(-\frac{1}{2}\frac{\gamma^0\gamma^j}{r^{2\lambda+1}}\textbf{M}+\frac{\gamma^0\gamma^ix_i}{r^{2\lambda+2}}\textbf{N}\right)\,,
\end{equation}

\noindent 
in terms of the $\textbf{M}$ and $\textbf{N}$ matrices, given as follows,
\begin{equation}
\textbf{M}=e^{i\ell\varphi}\left[
\begin{matrix}\frac{\sqrt{E-m}}{(2\pi)2E}\bigg(-i\ell\epsilon_{jk}\frac{x_k}{r}J_{|\ell-\alpha|}(kr)+kx_jJ'_{|\ell-\alpha|}(kr)\bigg)\\
\frac{\sqrt{E+m}}{(2\pi)2E}e^{i\varphi}\epsilon(\ell-\alpha)\bigg(-i(\ell+1)\epsilon_{jk}\frac{x_k}{r}J_{|\ell-\alpha|+\epsilon(\ell-\alpha)}(kr)+kx_jJ'_{|\ell-\alpha|+\epsilon(\ell-\alpha)}(kr)\bigg)\end{matrix}
\right]
\end{equation}

\noindent 
and
\begin{equation}
\textbf{N}=e^{i\ell\varphi}\left[
\begin{matrix}\frac{\sqrt{E-m}}{(2\pi)2E}krJ'_{|\ell-\alpha|}(kr)\\
\frac{\sqrt{E+m}}{(2\pi)2E}e^{i\varphi}\epsilon(\ell-\alpha)krJ'_{|\ell-\alpha|+\epsilon(\ell-\alpha)}(kr)\end{matrix}
\right]\,,
\end{equation}

\noindent
where the prime denotes derivation with respect to the argument of the Bessel functions.

After using the explicit form for the $\gamma$ matrices, one arrives at
\begin{eqnarray} \label{PsiHintNCPsiAntesdaDelta}
(\psi_{\hat k}^{(1)})^\dagger  H_{{\rm int},\lambda}^{(2)}  \psi_{k}^{(1)}  & = & - \theta \frac{e\Phi}{(4\pi)^3} \frac{k}{2E} \, \sum_{\hat \ell} \sum_{\ell} C_{\hat \ell}^{\dagger} C_{\ell} 
\nonumber\\
&\times& \Bigg\{e^{i(\ell-\hat \ell)\varphi+i {\hat \ell}\Omega} \epsilon(\ell-\alpha)\Bigg[\frac{\ell+1}{r^{2\lambda+1}} J_{|{\hat \ell} - \alpha|}({\hat k}r) J_{|\ell-\alpha|+\epsilon(\ell-\alpha)}(kr)
\nonumber\\
&+&\frac{k}{r^{2\lambda}}J_{|{\hat \ell}-\alpha|}({\hat k}r) J'_{|\ell-\alpha|+\epsilon(\ell-\alpha)}(kr)\Bigg]
\nonumber\\
&+&e^{i(\ell-{\hat \ell})\varphi+i({\hat \ell}+1)\Omega}\epsilon({\hat \ell}-\alpha)\Bigg[\frac{\ell}{r^{2\lambda+1}}J_{|{\hat \ell}-\alpha|+\epsilon({\hat \ell}-\alpha)}({\hat k}r)J_{|\ell-\alpha|}(kr)
\nonumber\\
&-&\frac{k}{r^{2\lambda}}J_{|{\hat \ell}-\alpha|+\epsilon({\hat \ell}-\alpha)}({\hat k}r) J'_{|\ell-\alpha|}(kr)
\nonumber\\
&-&2e^{i(\ell-{\hat \ell}) \varphi + i{\hat \ell} \Omega} \frac{k\epsilon(\ell-\alpha)}{r^{2\lambda}}J_{|{\hat \ell}-\alpha|}({\hat k}r)J'_{|\ell-\alpha|+\epsilon(\ell-\alpha)}(kr)
\nonumber\\
&+&2e^{i(\ell-{\hat \ell})\varphi+i({\hat \ell}+1)\Omega}\frac{k\epsilon({\hat \ell}-\alpha)}{r^{2\lambda}}J_{|{\hat \ell}-\alpha|+\epsilon({\hat \ell}-\alpha)}({\hat k}r)J'_{|\ell-\alpha|}(kr)\Bigg]\Bigg\} \,,
\end{eqnarray}

\noindent 
where $\varphi={\hat \varphi}+\Omega$, $\Omega$ being the scattering angle. The result in Eq.~(\ref{PsiHintNCPsiAntesdaDelta}) can be used in Eq.~(\ref{Born1}) to calculate the correction induced by the noncommutativity to the AB scattering amplitude, 
\begin{eqnarray} \label{Sfi 2pideltaFfi} 
T^{(2)}_{fi}&=&- i \lim_{\lambda\rightarrow1} \int dt \, \left\langle \psi^{(1)}_{f,\lambda} | H_{{\rm int},\lambda}^{(2)} | \psi^{(1)}_{i,\lambda} \right\rangle \,\nonumber\\ 
 &=& -i \lim_{\lambda\rightarrow1} \int dt\,d^2x \,\, 
\psi^{(1)}_{f,\lambda} (\vec{r}) \, H_{{\rm int},\lambda}^{(2)} \, \psi^{(1)}_{i,\lambda}(\vec{r}) \,\exp\left[{i(E_f-E_i)t}\right]
\nonumber
\\
&=&-2\pi i \, \delta(E_f-E_i)\,f^{(2)}_{fi}\,
\end{eqnarray}

\noindent 
where, 
\begin{equation}
f^{(2)}_{fi}=\,\lim_{\lambda \rightarrow 1} \, \int_{0}^{\infty}\int_{0}^{2\pi}rdrd\varphi\,\,
\psi^{(1)}_{f,\lambda}({r},\varphi) \, H_{{\rm int},\lambda}^{(2)} \, \psi^{(1)}_{i,\lambda}({r},\varphi) \,.
\end{equation}

\noindent
Using conservation of energy we can integrate in $\varphi$, thus obtaining,
\begin{eqnarray} \label{Ffi1}
f_{fi}^{(2)}&=&\theta\frac{e\Phi}{2(4\pi^2)}\frac{k}{2E}\lim_{\lambda\rightarrow1}\sum_{\ell}|C_{\ell}|^{2}
\nonumber
\\
&\times&\int_0^{\infty}dr\Bigg\{e^{i\ell\Omega}\epsilon(\ell-\alpha)\Bigg[\frac{-(\ell+1)}{r^{2\lambda}}J_{|\ell-\alpha|}(kr)J_{|\ell-\alpha|+\epsilon(\ell-\alpha)}(kr)
\nonumber 
\\
&+&\frac{k}{2r^{2\lambda-1}}J_{|\ell-\alpha|}(kr)J_{|\ell-\alpha|+\epsilon(\ell-\alpha)-1}(kr)-\frac{k}{2r^{2\lambda-1}}J_{|\ell-\alpha|}(kr)J_{|\ell-\alpha|+\epsilon(\ell-\alpha)+1}(kr)\Bigg]
\nonumber
\\
&+&e^{i(\ell+1)\Omega}\epsilon(\ell-\alpha)\Bigg[\frac{-\ell}{r^{2\lambda}}J_{|\ell-\alpha|+\epsilon(\ell-\alpha)}(kr)J_{|\ell-\alpha|}(kr)
\nonumber
\\
&-&\frac{k}{2r^{2\lambda-1}}J_{|\ell-\alpha|+\epsilon(\ell-\alpha)}(kr)J_{|\ell-\alpha|-1}(kr)
\nonumber
\\
&+&\frac{k}{2r^{2\lambda-1}}J_{|\ell-\alpha|+\epsilon(\ell-\alpha)}(kr)J_{|\ell-\alpha|+1}(kr)\Bigg]\Bigg\}\,.
\end{eqnarray}

The $r$ integrals can be performed using the analytical continuation of the formula
\begin{equation}
\label{formula}
\int_0^\infty dr\, \frac{J_\nu (kr) J_\mu (kr)}{r^\sigma} \,=\,\frac{2^{-\sigma} k^{\sigma -1} \Gamma(\sigma)\Gamma\left(\frac{1}{2} (\mu + \nu - \sigma + 1) \right)}{ \Gamma\left(\frac{1}{2} (\mu - \nu + \sigma + 1) \right) \Gamma\left(\frac{1}{2} (\mu + \nu + \sigma + 1) \right) \Gamma\left(\frac{1}{2} -\mu +\nu +\sigma + 1 \right) }\,,
\end{equation}

\noindent
originally valid for ${\rm Re} \, (\mu + \nu + 1) > {\rm Re} \, (\sigma) > 0$~\cite{watson}. At this point, it becomes clear what failed when we used a perturbation theory around solutions of the free Dirac equation: working out explicitly the expression in Eq.~(\ref{SfiHintNC}), one finds a factor similar to Eq.~(\ref{Ffi1}), but with Bessel functions of integer order $|\ell|$ and $|\ell|+\epsilon(\ell)$. Due to the asymptotic behavior of Bessel functions near the origin, $J_\alpha (x) \sim x^\alpha$, we cannot avoid the poles of the integral in Eq.~(\ref{formula}) for certain $\ell$, when $\lambda \rightarrow 1$. On the other hand, when the perturbation is performed starting with the solutions of the commutative Aharonov-Bohm problem the fact that $0<\alpha<1$ ensures that the order of the Bessel functions in Eq.~(\ref{Ffi1}) is never an integer and, in this way, by analytical continuation we avoid the poles of the $r-$integral.

It is convenient to separate the sum in $\ell$ in Eq.~(\ref{Ffi1}) in two parts, that with $\ell \le 0$ and that with
 $\ell \ge 1$, denoted respectively by $f^{(-)}_{fi}$ and $f^{(+)}_{fi}$. Using Eq.~(\ref{formula}), their result is,

\begin{eqnarray} \label{Ffi(-)}
f^{(-)}_{fi}&=&\theta\frac{e\Phi}{2(4\pi^2)}\frac{k}{2E}\sum_{\ell\le 0}\frac{k\alpha\left(e^{i\ell\Omega}+e^{i(\ell+1)\Omega}\right)}{4(\ell-\alpha)(\ell-\alpha+1)}\,,
\end{eqnarray}

\noindent
and
\begin{eqnarray} \label{Ffi(+)}
f^{(+)}_{fi}&=&\theta\frac{e\Phi}{2(4\pi^2)}\frac{k}{2E}\sum_{\ell\ge 1}\frac{-k\alpha\left(e^{i\ell\Omega}+e^{i(\ell+1)\Omega}\right)}{4(\ell-\alpha)(\ell-\alpha+1)}\,,
\end{eqnarray}

\noindent
so that, remembering that we are in a case where $\alpha \ll 1$,
\begin{equation} \label{SfiTotal}
T^{(2)}_{fi}=2\pi i\delta(E_f-E_i)\frac{\theta e\Phi}{2(4\pi^2)}\frac{k^2\alpha}{2E}\sum_{\ell=-\infty}^{\infty}\frac{\epsilon(\ell-\alpha)\left(e^{i\ell\Omega}+e^{i(\ell+1)\Omega}\right)}{4(\ell-\alpha)(\ell-\alpha+1)}\,.
\end{equation}

The dominant terms in the sum in Eq.~(\ref{SfiTotal}) are those for $\ell = 0$ and $\ell = -1$. It is convenient to separate these terms from the rest of the sum as follows,
\begin{eqnarray} \label{SfiFinal1} 
T^{(2)}_{fi}&=&\frac{\theta\pi
i}{2}\delta(E_f-E_i)\frac{e\Phi}{2(4\pi^2)}\frac{k^2\alpha}{2E}\Bigg\{\sum_{\ell=-1}^{0}\frac{(-1)(e^{i\ell\Omega}+e^{i(\ell+1)\Omega})}{(\ell-\alpha)(\ell-\alpha+1)}
\nonumber
\\
&+&\sum_{\ell \neq 0,-1}\frac{\epsilon(\ell-\alpha)e^{i\ell\Omega}}{\ell(\ell+1)}+\sum_{\ell \neq 0,-1}\frac{\epsilon(\ell-\alpha)e^{i(\ell+1)\Omega}}{\ell(\ell+1)}\Bigg\}\,,
\end{eqnarray}

\noindent
which, after some rearrangements can be cast as 
\begin{eqnarray} \label{SfiFinal2} 
T^{(2)}_{fi}&=&\frac{\theta\pi
i}{2}\delta(E_f-E_i)\frac{e\Phi}{2(4\pi^2)}\frac{k^2}{E}\Bigg\{\alpha+i\sin\Omega+\alpha\cos\Omega+\frac{i\alpha\sin\Omega}{2}
\nonumber
\\
&+&\frac{4i\alpha}{2}\sum_{\ell=2}^{\infty}\frac{\sin(\ell\Omega)}{\ell^2-1}\Bigg\}
\nonumber
\\
&=&\frac{\theta\pi
i}{4(4\pi^2)}\delta(E_f-E_i)\frac{k^2}{E}\Bigg\{2\pi i\alpha\sin\Omega+\alpha^2\left[1+e^{i\Omega}-2i\ln\left(2\sin\frac{\Omega}{2}\right)\right]\Bigg\}\,
\end{eqnarray}

\noindent 
(it is necessary to remember that $\alpha$ is small to reorganize the sums and obtain the exact result in Eq.~(\ref{SfiFinal2})).

Finally, one can use the result in Eq.~(\ref{SfiFinal2}), together with the one in Eq.~(\ref{Born1}), to obtain the transition probability as follows,
\begin{equation} \label{ModA}
|T_{fi}|^2=|T^{(1)}_{fi}|^2 + 2 \,{\rm Re} \left[  T^{(1)}_{fi} \, (T^{(2)}_{fi})^\dagger \right] \,.
\end{equation}

\noindent 
Since we are considering a small flux limit, the ``unperturbed'' amplitude $T^{(1)}_{fi} = \left\langle \, \psi^{(1)}_f \,|\,\psi^{(1)}_i \, \right\rangle$ can itself be calculated in a Born approximation, thus obtaining for the differential cross section of a spin one-half particle in the noncommutative AB potential
\begin{equation} \label{difcross}
\frac{d\sigma}{d\Omega}=\int \frac{1}{2\pi}\Big(\frac{\alpha}{2E}\Big)^2 \frac{1}{k/E}
 \left[\frac{1}{\sin^2(\Omega/2)}-\theta
  k^2 \cos^2(\Omega/2)\right]\delta(E_k-E_{\hat k}) \, {\hat k} d{\hat k} \,.
\end{equation}

The expression in Eq.~(\ref{difcross}) is the usual (commutative) Aharonov-Bohm differential cross-section, corrected by a $\theta$ dependent contribution arising from the noncommutativity, which we calculated to the first order in the small parameter $\theta$. This correction, integrated over energies, reads
\begin{eqnarray} 
\delta \left( \frac{d\sigma}{d\Omega}\right) \,=\,-\frac{\theta}{2\pi}\frac{\alpha^2}{4}k\cos^2\frac{\Omega}{2}\,,
\end{eqnarray}

\noindent
therefore the modification induced by the noncommutativity to the commutative cross-section is
\begin{eqnarray} \label{SChTotal}
\delta \left( \sigma \right) &=&-\frac{\theta\alpha^2 k}{8}\,.
\end{eqnarray}

\noindent
This correction vanishes in the $\theta \rightarrow 0$ limit so that no singularities are generated in taking the commutative limit.

Phenomenological bounds limit the noncommutativity parameter $\theta$ to be at the most of order $\sim (10 {\rm Tev})^{-2}$~\cite{Chaichian:2000si,Carroll:2001ws,Mendez18}. Looking at Eq.~(\ref{difcross}), it becomes clear that the effects of noncommutativity are negligible except for particles of extremely high energy. 

In this work we studied the corrections to the scattering of relativistic spin $1/2$ particles induced by the noncommutativity of spacetime, in the small flux approximation. We adopted a distorted Born approximation to calculate the transition amplitudes, considering the $\theta$ dependent part of the potential as a perturbation over the exact solution of a spin $1/2$ particle in a commutative AB potential. In this way, we avoided singularities due to the behavior of the noncommutative potential near the origin, and we were able to calculate the corrections induced by the noncommutativity to the differential and total cross sections. No singularities are found by taking $\theta \rightarrow 0$, so that the commutative limit of this model is regular. It is interesting to contrast this behavior from the one in a quantum field theory context~\cite{Anacleto:2004xs}. When considering a Dirac fermion coupled to the noncommutative Chern-Simons field, an appropriately adjusted Pauli term is need to avoid infrared singularities that are generated from the so-called UV/IR mixing~\cite{Minwalla5}. At the level of approximation we are working with, for the case of a fermion minimally coupled to the AB potential, no singularities were found in the commutative limit.

\vspace{1cm}

\textbf{Acknowledgments}

This work was partially supported by the Brazilian agencies Funda\c{c}\~{a}o de Amparo 
\`{a} Pesquisa do Estado de S\~{a}o Paulo (FAPESP) and Conselho 
Nacional de Desenvolvimento Cient\'{\i}fico e Tecnol\'{o}gico (CNPq). The work of A. F. F. was supported by FAPESP,  project 04/13314-4.

\end{document}